# Further evidence on the effect of magnetism on lattice vibrations: the case study of sigma-phase $Fe_{0.525}Cr_{0.455}Ni_{0.020}$ alloy

Stanisław M. Dubiel[1*] and Jan Żukrowski[2]

[1]Faculty of Physics and Applied Computer Science, [2]Academic Center for Materials and Nanotechnology, AGH University of Science and Technology,
PL-30-059 Kraków, Poland

**Abstract**

Sigma-phase $Fe_{0.525}Cr_{0.455}Ni_{0.020}$ alloy was studied by means of Mössbauer spectrometry in the temperature range of 5-293 K. The average center shift, <*CS*>, determined from the recorded Mössbauer spectra was shown to significantly deviate from the Debye model in the temperature range below ~50 K i.e. in the magnetic state of the sample. The deviation is obviously due to the effect of magnetism on the lattice dynamics of the studied sample. Analysis of the <*CS*>-data in the paramagnetic phase in terms of the Debye model yielded the value of 437(7) K for the Debye temperature.

* Corresponding author: Stanislaw.Dubiel@fis.agh.edu.pl

# 1. Introduction

The role of magnetism in the lattice dynamics is usually considered as negligible. Such belief seems to follow from calculations according to which the spin susceptibility of metal is not affected by electron-phonon interactions (EPI) ([1] and references therein). The effect of the EPI was estimated as $\hbar\omega_D/\varepsilon_F \approx 10^{-2}$ ([1] and references therein) where $\varepsilon_F$ is the Fermi energy, and $\hbar\omega_D$ is the Debye energy. However, Kim showed [2] that the impact of the EPI on a spin susceptibility can be strongly, viz. by two orders of magnitude, enhanced by exchange interactions between electrons. In other words, the consequence of the EPI on magnetic properties of metallic systems, and *vice versa*, is much more significant than generally believed. From experimental viewpoint good candidates for testing these predictions are magnetic systems in which the magnetism is highly delocalized (itinerant). To select good candidates for testing the Kim's prediction one can use the Rhodes-Wohlfarth criterion (plot) according to which the magnetic system are itinerant if the Rhodes-Wohlfarth ratio is higher than 1 [3].

Sigma-phase Fe-based compounds e. g. Fe-Cr, Fe-V, Fe-Mo or Fe-Re, to name just binary ones, seem to be particularly well-suited to study the effect of magnetism on the lattice dynamics as, following the Rhodes-Wohlfarth criterion, their magnetism is highly itinerant [4,5]. Indeed, our previous Mössbauer spectrometric studies on σ-FeX compounds (X=Cr, V) gave evidence that both spectroscopic parameters relevant to the lattice vibrations, viz. the center shift as well as the recoil-free factor, showed anomalous behavior in the paramagnetic and magnetically ordered states [6,7]. Noteworthy, also the applied magnetic field significantly affected the lattice dynamics [7]. The same experimental technique also revealed strong anomalies in the center shift and in the electric field gradient in a C14 Lave phase $Nb_{0.975}Fe_{2.025}$ compound on a transition from a paramagnetic to ferromagnetic state [8]. Recently, using the Mössbauer effect on $^{119}$Sn nuclei diffused into chromium (highly itinerant antiferromagnet) an anomaly in the behavior of the center shift was revealed at the Néel temperature, $T_N$ [9]. In this case an increase of the kinetic energy of $^{119}$Sn atoms vibrations in the temperature range of 3.5 K below $T_N$ amounted to 8.5 meV from which -0.27 meV can be accounted for by the thermal effect.

Furthermore, the effect of magnetism on the lattice dynamics can be experimentally studied by measuring a sound velocity, $v_s$, or/and a phonon frequency of a magnetic system above and below the magnetic ordering temperature, $T_C$ [10]. Alternatively, one can study the effect of an external magnetic field, $H$, on $v_s$. The predicted change in $v_s$ is proportional to $H^2$ for $T > T_C$, and to $H$ for $T < T_C$ [10]. Using the nuclear inelastic scattering technique, a strong increase of the sound velocity was revealed in the magnetic state relative to its value in the paramagnetic phase of σ-FeCr alloys [11].

In this paper we report further evidence on the effect of magnetism on the lattice vibrations as found by means of the [57]Fe-site Mössbauer spectrometric study performed vs. temperature on a σ-Fe$_{0.525}$Cr$_{0.455}$Ni$_{0.020}$ alloy.

## 2. Experimental

### 2.1. Sample preparation and characterization

The sigma-phase sample of Fe$_{0.525}$Cr$_{0.455}$Ni$_{0.020}$ was prepared from a bcc-Fe-Cr-Ni master alloy having the corresponding chemical composition. The homogenized master alloy was isothermally annealed at 973 K for 7 days under a dynamic vacuum. The verification of the alpha-to-sigma phase transformation was checked by X-ray and neutron diffraction techniques. More details on this aspect of the study can be found elsewhere [12].

### 2.2. Measurements

[57]Fe Mössbauer spectra were recorded in a transmission mode by means of a standard spectrometer (Wissel GmbH) equipped with a drive operating in a sinusoidal mode. The 14.4 keV gamma rays were emitted by a [57]Co/Rh source. Its activity made possible recording a spectrum of a very good statistical quality within a 1-2 days run. The spectra were registered in a 1024-channel analyzer in the temperature ($T$) range of 5-293 K subdivided into two areas: (a) 5-100 K and (b) 90-293 K. A closed-cycle Janis Research 850-5 Mössbauer Refrigerator System was used for recording the spectra in the (a)-area whereas for the (b)-area a standard Janis SVT-200 cryostat was employed. The temperature was kept constant to the accuracy better than ±0.2 K and ±0.1 K for the former and the latter cryostat, respectively. Examples of the measured spectra are presented in Fig. 1.

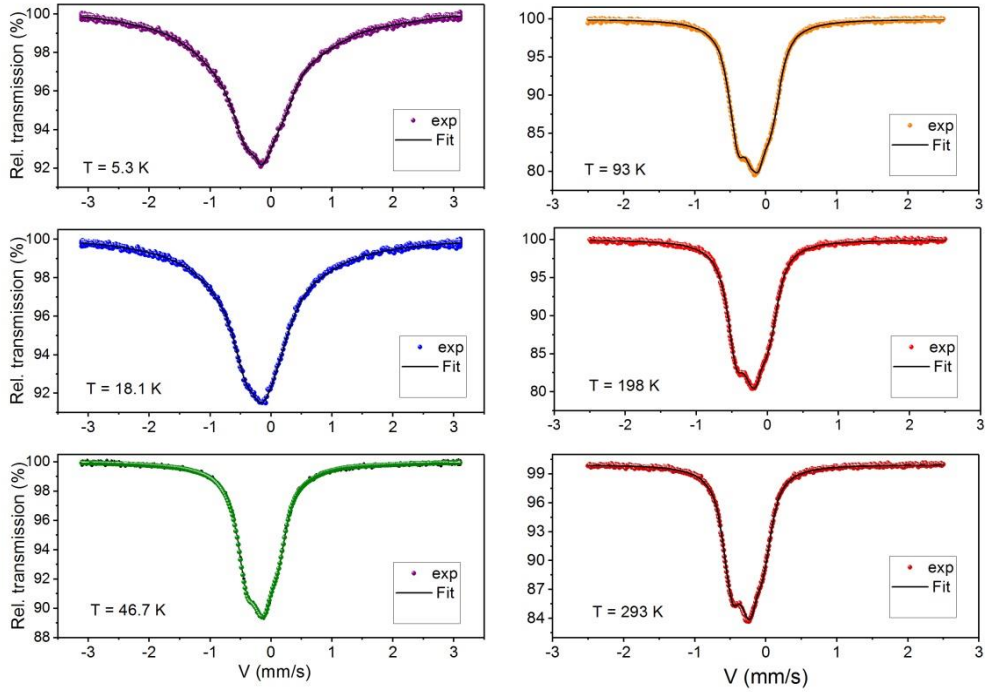

Fig. 1 Examples of the spectra recorded for the studied sample at various temperatures shown.

## 3. Results and discussion

### 3.1. Spectra analysis

The spectral parameter relevant to the lattice dynamics is the center shift of the spectra, *CS*, so the aim of the spectra analysis was to determine its value precisely. The analysis of the presently measured spectra is not trivial, as they are poor resolved and there are five non-equivalent lattice sites in the unit cell of the sigma-phase. To achieve our goal two different fitting procedures were used. First, the spectra were fitted to a Gaussian profile. It permitted to obtain the average center shift, <*CS*>, (position of the center of the Gaussian peak) and the width of the spectrum at half absorption maximum (FWHM=2.35·σ, where σ stands for the standard deviation). This procedure has turned out to yield good values of both parameters. In particular, the obtained σ-values reproduce very well the widths of the spectra at half absorption maximum – see Fig. 2. Additionally, the spectra measured in the magnetic phase ($T \leq \sim 50$ K [13]) were fitted to a single line having the Voigt profile. Determined therefrom <CS>-values were within the error limit the same as the ones found with the Gaussian profile.

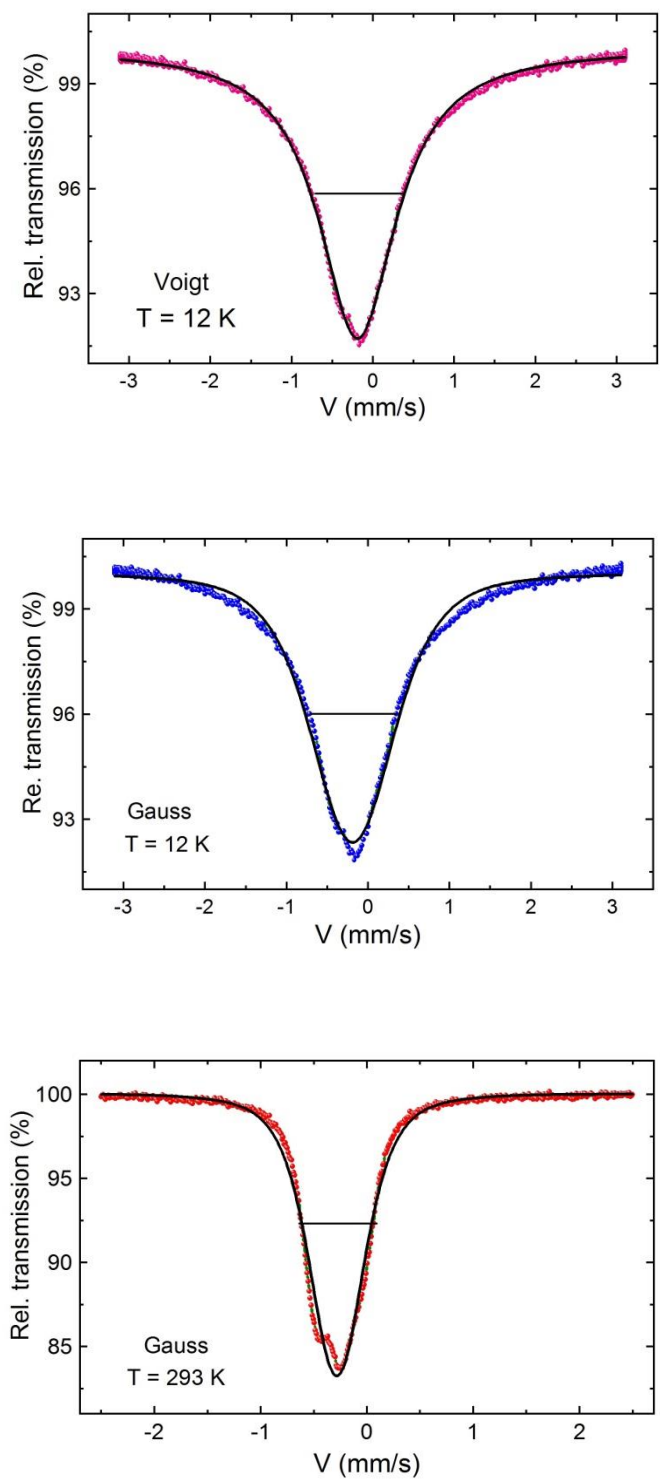

Fig. 2 Examples of the spectra fitted to Gaussian and Voigt profiles. The horizontal bars stand for the full width of the spectra at half maximum.

The spectra in the paramagnetic phase i.e. those recorded at $T \geq 50$ K were analyzed in terms of five doublets, each corresponding to one of five different lattice sites present in a unit cell of the sigma-phase. The fitting procedure was based on a transmission integral method. Free parameters in the analysis of the non-magnetic spectra were: center shift, $CS_k$ (k=1, 2, 3, 4, 5), main component of the electric field gradient, $V_{zzk}$, and line width, $\Gamma_k$, while the relative abundance of the components was kept constant in line with the results obtained from the neutron diffraction investigation [1]. The contribution weighted values of the center shift perfectly agree with the corresponding ones obtained from the Gaussian approach – see Fig. 3.

### 3.2. Debye temperature

The knowledge of the Debye temperature, $T_D$, is regarded in solid state physics as an important parameter, because it is pertinent to various physical properties of materials, such as specific heat, elastic constants, melting point and others. Experimentally $T_D$ can be determined by means of different techniques including Mössbauer spectrometry. With the latter, $T_D$ can be figured out either from a temperature dependence of a center shift, $CS$, or that from a recoil-free fraction. The temperature dependence of the center shift, $CS(T)$, reads as follows:

$$CS(T) = IS(T) + SOD(T) \qquad (1)$$

Where $IS$ stays for the isomer shift and $SOD$ is the so-called second order Doppler shift i.e. a quantity related to a non-zero mean value of the square velocity of vibrations, $<v^2>$, hence a kinetic energy. Following the Debye model for lattice vibrations, and assuming that $IS$ hardly depends on temperature, so it can be ignored [14], the temperature dependence of $CS$ is practically related to the second term in eq. (1) which is related to $T_D$ via the following relationship [15]:

$$CS(T) = IS(0) - \frac{3k_B T}{2mc}\left(\frac{3T_D}{8T} + 3\left(\frac{T}{T_D}\right)^3 \int_0^{T_D/T} \frac{x^3}{e^x - 1} dx\right) \qquad (2)$$

Here $m$ stays for the mass of the $^{57}$Fe atom, $k_B$ is the Boltzmann constant, $c$ is the velocity of light, and $x = \frac{\hbar \omega}{2\pi k_B T}$ ($\omega$ being frequency of vibrations).

Figure 3 illustrates temperature dependence of experimentally found average values of the center shift, <CS>(T), using both fitting procedures. Clearly, a significant deviation of the data from the Debye model occurs for T<~50 K. Consequently, the anomalous data points were not taken into account in the fitting of the <CS>-data to the eq. (2). The best-fit is shown as a solid line, and the resulted value of $T_D$ = 437(7) K. Noteworthy, this value of $T_D$ is within the error limit the same as the one found for the sigma-phase $Fe_{54}Cr_{46}$ alloy [16]. It should be at this point emphasized that similar anomalies in <CS>(T) were previously observed for other Frank-Kasper phases like binary Fe-based sigma-phase alloys [6,7] and for the C14 Laves phase (known also as λ- $Nb_{0.975}Fe_{2.025}$ compound [8]. The common feature of all these samples is an itinerant character of their magnetism.

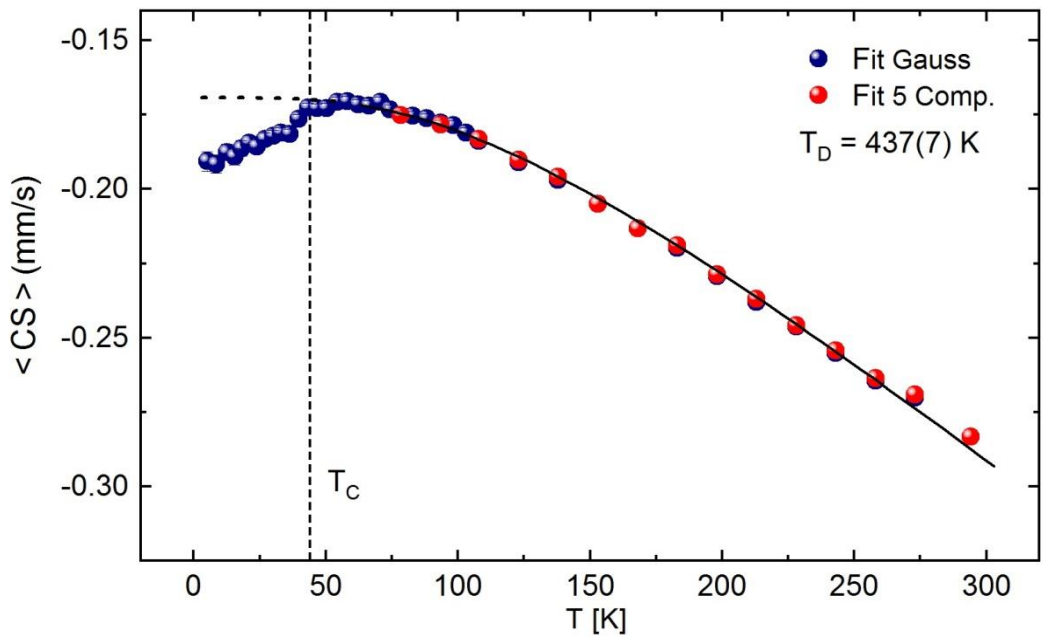

Fig. 3 Temperature dependence of the average center shift, <CS>. The solid line stays for the best fit of the data to eq. (2). The determined therefrom Debye temperature, $T_D$, is shown. The vertical dashed line indicates the magnetic ordering temperature, $T_C$, found with magnetization measurements [13].

Concerning the behavior of *<CS>* in the magnetic state ($T <$ ~50 K), its decrease is observed on lowering temperature. The decrease (increase of negative values) indicates an increase of the average square-velocity of $^{57}$Fe atoms vibration in the lattice, $<v^2>$, hence and increase of the related kinetic energy, $E_k$. The latter sounds paradoxically, as from the thermodynamic viewpoint $E_k$ should be decreasing with decreasing $T$. The opposite behavior means that the increase of $E_k$ has its source in the magnetism. The observed maximum increase of $<v^2>$ amounts to ~2·10$^4$ (m/s)$^2$ whereas that of $E_k$ to ~6 meV. This is significantly more than the corresponding values determined for the σ-Fe$_{0.52}$Cr$_{0.48}$ sample [6]. The difference may be due to the presence of Ni in the presently investigated sample.

**3.3. Mössbauer spectrometric indication of magnetism**

The temperature at which the drop of *<CS>* begins correlates quite well with the value of $T_C$ determined by the magnetization measurements [13]. It is, however, of interest to see how the temperature of the anomaly in *<CS>(T)* compares to the temperature at which the Mössbauer spectra start to broaden. The latter can be taken as an indication of a transition into the magnetic state. The line width can be taken as a good measure of spectral broadening. Figure 4 illustrates the standard deviation of the Gaussian profiles fitted to the spectra as a function of $T$. It is clear that a sharp increase of the standard deviation takes place for $T \leq$~55 K, hence ~10 K higher than $T_C$ reported based on magnetization measurements [13]. The difference can be understood in terms of two factors viz. the differences in the time window between the two techniques (~10$^{-7}$s for the Mössbauer effect and ~10$^2$s for the magnetization) and that in the sensitivity to spin correlations in the real space (few lattice spacing for the the Mössbauer effect and few mm for the magnetization).

A light increase of the standard deviation seen between ~55 K and ~110 K is likely due to a slight line broadening of the spectra recorded in the closed-cycle Janis Research 850-5 Mössbauer Refrigerator System.

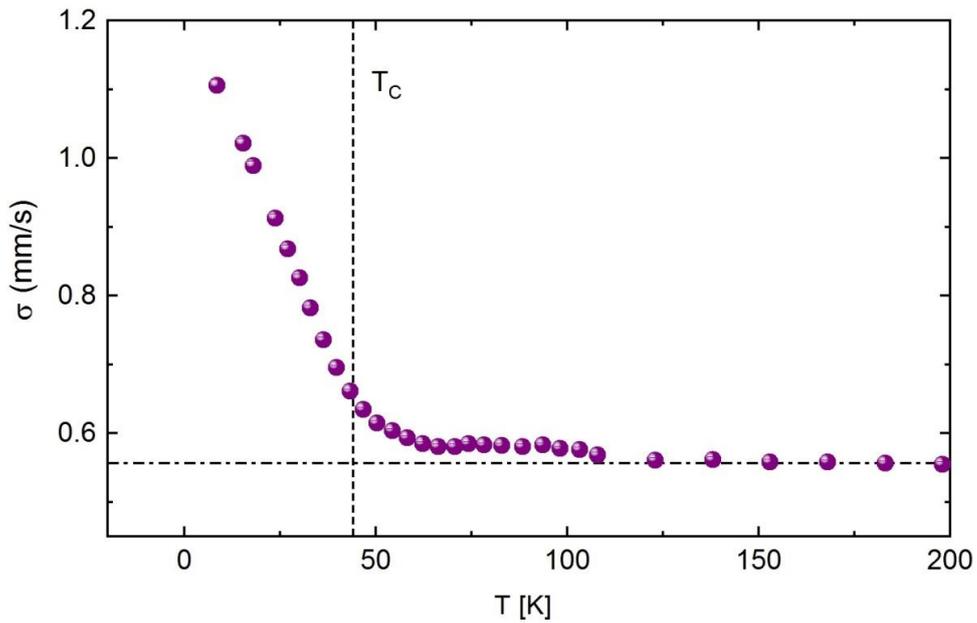

Fig. 4 Standard deviation (σ) of the Gaussian profile fitted to the spectra vs. temperature. The vertical line marks the position of the Curie temperature as revealed from magnetization measurements [13]. The horizontal line indicates values of the standard deviation found from the spectra measured in the Janis SVT-200 cryostat.

## 4. Conclusions

Based on the results obtained in this investigation, the following conclusions can be drawn:

1. The temperature dependence of the average isomer shift of the studied sigma-phase $Fe_{0.525}Cr_{0.455}Ni_{0.020}$ alloy, <CS>, exhibits a significant deviation from the behavior predicted by the Debye model in the temperature range below ~50 K.

2. The temperature at which this anomaly begins coincides well with the magnetic ordering (Curie) temperature, so the anomaly can be regarded as due to effect of magnetism on lattice vibrations.

3. The maximum decrease of <CS> taking place at ~5K corresponds to an increase of the average square velocity of $^{57}$Fe atoms vibrations by ~2·10$^4$ (m/s)$^2$ equivalent to an increase of the corresponding kinetic energy by ~6 meV.

4. The value of the Debye temperature of 437(7) K was determined from the temperature dependence of <*CS*> in the paramagnetic range.

**Acknowledgements**

This work was financed by the Faculty of Physics and Applied Computer Science AGH UST and ACMIN AGH UST statutory tasks within subsidy of Ministry of Science and Higher Education, Warszawa.

**CRediT author statement**

**SMD:** Conceptualization, Validation, Formal analysis, Resources, Writing - Original Draft, Visualization: **JŻ:** Software, Validation, Formal analysis, Investigation, Data Curation.